\documentclass[article,twocolumn]{revtex4}       
\usepackage{latexsym}
\usepackage{graphicx}

\usepackage{amssymb}   
\usepackage{epsfig}
\usepackage{amsmath}

\begin{document}

\title{Magnon Dispersion and Single Hole Motion in 2D Frustrated Antiferromagnets with Four-Sublattice Structures} 

\author{Satyaki Kar}


\affiliation{Saha Institute of Nuclear Physics, Salt Lake, Kolkata 700064, India.}
\begin{abstract}
We study a two dimensional spin-$\frac{1}{2}$ $J_1-J_2$ antiferromagnet in a square lattice using the linearized spin wave theory recognizing the 4-sublattice nature of the underlying magnetic lattice. Multiple magnon modes with optical and acoustic branches about the stable Neel ordered and double acoustic branches about the columnar reference states are obtained for small and large values of $\lambda(=J_2/J_1)$ respectively. An additional uniaxial anisotropy, for large $\lambda$, can lead to distinct spin gaps in such systems, as also witnessed experimentally. The single hole spectral behavior in a 2D $t-J_1-J_2$ model, for small frustration, is then calculated within the non-crossing approximation. Our results match fairly well with exact diagonalization results from a $4\times4$ cluster. Hole spectral features and their evolution with $\lambda$ resulting in ``water-fall''-like smooth spectral weight transfer are discussed. Hole energy bands are identified and the corresponding energy-shift and reduction in width with spin-frustration are indicated. 
\end{abstract}


\maketitle   

\section{Introduction}
Geometrically frustrated quantum Heisenberg antiferromagnet (AF) in low dimensions have long emerged as a popular field of study in condensed matter in understanding the magnetic properties of various magnetic materials with layered structures. As for example, the key to charge transport in the newly discovered Iron-based high temperature superconductors\cite{feas1,feas2} is believed to exist in their spin-frustrated conducting FeAs layers. Or, the spin-$\frac{1}{2}$ $V^{4+}$ ions in recently synthesized vanadium oxide compounds like $Li_2VO(Si,Ge)O_4$ are found to possess pairs of frustrating super-exchange paths among them in their two-dimensional (2D) magnetic layers\cite{melzi1,melzi2}.
\begin{figure}[htp]
\begin{center}
\epsfig{file=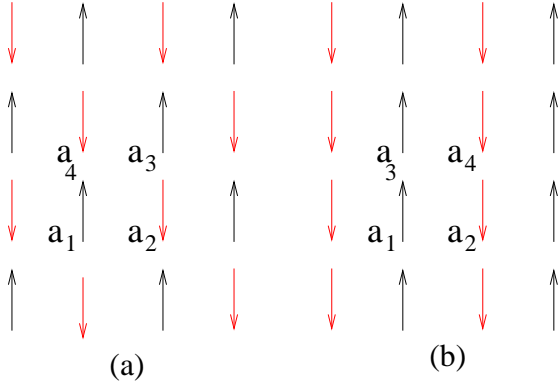,width=.85\linewidth,clip=}
\end{center}
\caption{The (a) Neel ordered $k=(\pi,\pi)$ reference state and (b) columnar $(\pi,0)$ reference state with 4 sublattices. Here $a_i$ denotes the $i$-th sublattice.}
\label{fig1}
\end{figure}
A 2D $J_1-J_2$ model, with competing $J_1$ and $J_2$ AF interactions, gives different phases in different parameter range of $\lambda=J_2/J_1$. A classically Neel-ordered phase with $Q=(\pi,\pi)$ magnetic long range order is formed for small $\lambda$ whereas a columnar $Q=(\pi,0)$ and/or $(0,\pi)$ ordered phase is obtained for large values of $\lambda$. Numerous works have been done on this model using different methods like exact diagonalization\cite{schulz,dagotto}, spin wave calculations\cite{trumper,sushkov}, series expansion\cite{oitmaa} etc. that resulted in these conclusions. Quantum fluctuations, on the other hand, result in a disordered spin liquid phase in such system for intermediate values of frustration: $0.4<\lambda<0.6$\cite{schulz,oitmaa,sorella}. Chandra et al.\cite{chandra} used conventional spin wave theory to study this model and predicted a spin-liquid ground state for large values of frustration. Later, Dagotto et al.\cite{dagotto} used exact diagonalization to show how magnetization changes with $\lambda$. Their calculations indicated Neel ordered AF phase and columnar phase at small and large values of $\lambda$ respectively. Recently, Li et al.\cite{sorella} used a bosonic resonating valence-bond ansatz to find out that stable spin liquid phases are indeed there in such systems for moderate values of frustration: $0.4<\lambda<0.6$.

The present work involves studying the spin ordered phases of a spin-$\frac{1}{2}$ 2D $J_1-J_2$ model using a four-sublattice description. We use the linear spin wave (LSW) approximation to identify the magnon modes about the $(\pi,\pi)$ and $(\pi,0)$ reference states for $\lambda<0.4$ and $\lambda>0.6$ respectively and consider spin wave expansion about these states. Our calculations are based on the four sublattice nature of the problem which comes naturally from the competing $J_1$ and $J_2$ interactions in the Hamiltonian. Our results derived are found compatible with the earlier two-sublattice calculations\cite{thalmeier,hamad,erica} on the model, because the magnetic reference states, in the two limits that we consider, retains their two-sublattice nature. We emphasize at this point that the competing interaction pair ensures the four sublattice nature to the problem and, as a consequence, the appearance of multiple magnon branches in the system whose signatures are witnessed experimentally, $e.g.$, in neutron diffraction results\cite{ewings,zhao,conceicao} in various frustrated magnetic systems. 

Our work also include a study of the $t-J_1-J_2$ model in an undoped frustrated antiferromagnet to witness the hole motion in it. Self consistent Born approximation (SCBA) is used to sum up the self energy diagrams within the non-crossing approximation and the single hole spectra are obtained for the system with the underlying four-sublattice structure. The main objective of this paper is to form a building block in understanding the magnon dispersions and the magnon-mediated hole-hole interactions in a 2D frustrated antiferromagnet that can be realized and later on compared with Neutron diffraction or angle-resolved photo-emission spectroscopy (ARPES) results obtained from various frustrated quasi-2D spin systems like $Li_2VO(Si,Ge)O_4$\cite{melzi1,melzi2} or the FeAs superconducting compounds\cite{xia,nakayama}.

The paper is organized as follows: Section 2 is the formulation part which describes a 2D $J_1-J_2$ spin Hamiltonian and the conventional linear spin wave analysis on it. In Section 3, we discuss the Bogoliubov transformation yielding the magnon modes for different values of $\lambda$. Section 4 deals with the hole hopping using a $t-J_1-J_2$ model. SCBA is used to obtain the hole spectral functions and the spectral behavior are discussed in detail. Finally in section 5, we summarize our results and discuss the importance of the work presented.

\section{Formulation}
In a 2D square lattice that we consider, a $J_1-J_2$ Heisenberg spin Hamiltonian is given as
\begin{align}
H_{J1J2}=H_{J1}+H_{J2}=J_1\sum_{<i,j>}{\bf{S_i}.{S_j}}+J_2\sum_{<<i,j>>}{\bf{S_i}.{S_j}}
\label{eq1}
\end{align}
where $<i,j>$ of $H_{J1}$ and $<<i,j>>$ of $H_{J2}$ represent indices for the nearest neighbor (NN) and next nearest neighbor (NNN) site-pairs respectively and $S$ denotes the spin. $H_{J1}$ produces two opposite spin sublattices and at low temperature, the elementary excitations of spin waves or magnons are created on top of this $(\pi,\pi)$ AF reference state.

$H_{J2}$ , however, involves NNN AF spin interactions that do not communicate between $\uparrow$ and $\downarrow$ sublattices caused by $H_{J_1}$. Rather within each such sublattice, it creates two sublattices on its own. Thus an overall outer product of four spin sublattices are formed in a $J_1-J_2$ antiferromagnet, as also demonstrated in Ref.\cite{sachdev} (see Fig.\ref{fig1}).

This $J_1-J_2$ model represents a frustrated spin system as the $J_1$ and $J_2$ terms prefer different spin-orders to be the ground state of the system. 
For comparatively small strength of $J_2$ with the factor $\lambda=J_2/J_1$ being a small fraction, the $(\pi,\pi)$ spin order describes the reference state of the system and the spin excitations appear above this vacuum state. At large enough value of $\lambda$ however, a $(\pi,0)$ or $(0,\pi)$ ordered state becomes the zero spin deviation vacuum state of the system.

Here in this work, we use LSW approximation\cite{stratos} to find the magnon modes of the $J_1-J_2$ model about $(\pi,\pi)$ and $(\pi,0)$ spin ordered states at small and large values of $\lambda$ respectively.

Figure \ref{fig1} illustrates the real space snapshot of the $(\pi,\pi)$ and $(\pi,0)$ reference states with the indication of the 4 sublattices. The unit vectors in real space for the $J_1-J_2$ model are $a1=2\hat{x}$ and $a2=2\hat{y}$ (and $b_1=\pi\hat{y}$ and $b_2=\pi\hat{x}$ in the Fourier space). The 1st Brillouin zone (BZ) is a square with corners at $(\pm\pi/2,\pm\pi/2)$. The lattice is broken down to 4 sublattices and the LSW calculation gives two different spin wave modes each with degeneracy 2 (for the spin up-down symmetry).

 The bosonic spin wave operators $a_{i,{\bf r}}$'s are defined as $S_{z,i}=S-a_{i,{\bf r}}^\dagger a_{i,{\bf r}}~(-S+a_{i,{\bf r}}^\dagger a_{i,{\bf r}})$ for the pair of $\uparrow~(\downarrow)$ sublattices (non-identical, because they experience non-zero interaction with each other) and they follow the Fourier transformation: $a_{i,{\bf k}}=\sqrt{\frac{4}{N}}\sum_{{\bf r}\epsilon i}a_{i,{\bf r}}~e^{i{\bf k.r}}$, ${\bf k}$ being the Bloch wave vectors within the BZ. This transforms our Eq.\ref{eq1} to $H=E_0+\sum_k<{\bf a_k}|A|{\bf a_k}>$ with $E_0$ being a constant term and the column vector $|{\bf a_k}>=(a_{1,{\bf k}}, a_{2,-{\bf k}}^\dagger, a_{3,{\bf k}}, a_{4,{-\bf k}}^\dagger)$. For small $\lambda$ with $(\pi,\pi)$ reference state, we find
\begin{displaymath}
A=4J_1S\left(\begin{array}{cccccccc}
1-\lambda & \frac{cos~ k_x}{2} & \lambda\Gamma_k & \frac{cos~ k_y}{2}\\
\frac{cos~ k_x}{2} & 1-\lambda & \frac{cos~ k_y}{2} & \lambda\Gamma_k\\
\lambda\Gamma_k & \frac{cos~k_y}{2} & 1-\lambda & \frac{cos~ k_x}{2}\\
\frac{cos~ k_y}{2} & \lambda\Gamma_k & \frac{cos~ k_x}{2}  & 1-\lambda
\end{array}\right)
\end{displaymath}\\
with $\Gamma_k=\frac{1}{2}($cos$(k_x+k_y)+$cos$(k_x-k_y))$ . On the other hand, for large $\lambda$ with $(\pi,0)$ reference state, we obtain 
\begin{displaymath}
A=4J_1S\left(\begin{array}{cccccccc}
\lambda & \frac{cos~ k_x}{2} & \frac{cos~ k_y}{2} & \lambda\Gamma_k\\
\frac{cos~ k_x}{2} & \lambda & \lambda\Gamma_k & \frac{cos~ k_y}{2} \\
\frac{cos~ k_y}{2} & \lambda\Gamma_k & \lambda & \frac{cos~k_x}{2}\\
\lambda\Gamma_k & \frac{cos~ k_y}{2} & \frac{cos~ k_x}{2} & \lambda~~~
\end{array}\right).
\end{displaymath}

 Here, it should be noted that in the $J_1-J_2$ Hamiltonian, the mere presence  of the $J_2$ term imposes four sublattices to the system irrespective of the strength of frustration $\lambda$. But because at small and large values of $\lambda$, where $J_1$ and $J_2$ term dominates the magnetics respectively, two-sublattice magnetic stable reference states can be found. And thus results from the four-sublattice calculations in those limits can also be obtained from mere two-sublattice considerations of the problem followed by folding of the bands as per the requirement of unit cell doubling in the $J_1-J_2$ Hamiltonian. But that can not refrain one from carrying out the calculations using the actual four-sublattice structure, as is done in this present paper, which is the general feature of the problem as far as the Hamiltonian is concerned.
\begin{figure}
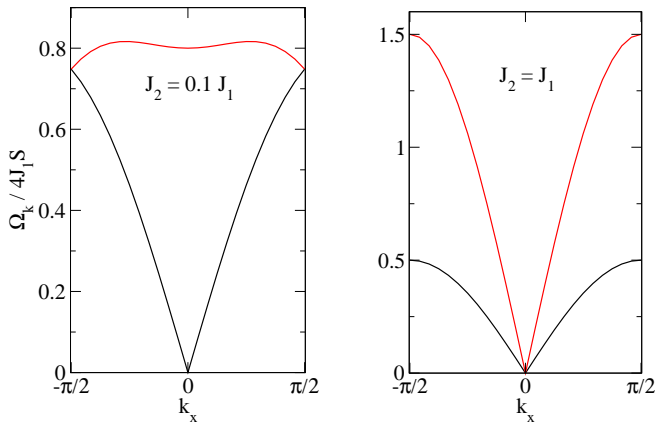

\begin{center}
\epsfig{file=figure2a.eps,width=.48\linewidth,clip=}\hskip .3 in
\epsfig{file=figure2b.eps,width=.43\linewidth,clip=}
\end{center}
\caption{Left:Acoustic (black) and optical (red or grey) magnon modes about $(\pi,\pi)$ reference state. Right: Acoustic magnon modes about $(\pi,0)$ reference state. Both the plots have $k_y=0$.}
\label{fig2}
\end{figure}
\section{Magnon Dispersion}
This Hamiltonian is diagonalized using Bogoliubov transformation which is a canonical transformation where new bosonic operators $\alpha_k$'s are introduced as 
\begin{align}
a_{i,{\bf k}}&=U_{i,jo}({\bf k})\alpha_{jo,{\bf k}}+U_{i,je}({\bf k})\alpha_{je,{-\bf k}}^\dagger,~~~(i~~odd)\nonumber\\
a_{i,-{\bf k}}^\dagger&=U_{i,jo}({\bf k})\alpha_{jo,{\bf k}}+U_{i,je}({\bf k})\alpha_{je,{-\bf k}}^\dagger,~~~(i~~even)
\label{eq2}
\end{align}
where sum over jo (1, 3) and je (2, 4) are implied, $U$ being the $4\times4$ coefficient matrix. The bosonization condition for the variables $\alpha$ requires
\begin{eqnarray}
\sum_j(-1)^{i+j}U_{i,j}({\bf k})U_{i^\prime,j}({\bf k})=\delta_{i,i^\prime}
\label{eq3}
\end{eqnarray}
while the diagonalization of the Hamiltonian matrix requires
\begin{eqnarray}
\sum_j(A_{ij}({\bf k})-\delta_{ij}\epsilon_l({\bf k})(-1)^{l+j})~U_{j,l}({\bf k})=0.
\label{eq4}
\end{eqnarray}
$\epsilon_l({\bf k})$ denoting the $l$-th eigen-energy mode.
Non-trivial solution of Eq.\ref{eq4} gives the eigenvalues.
Fig.\ref{fig2} shows the magnon modes for $k_y=0$ for both small and large $\lambda$'s.

For $\lambda<0.5$ with ($\pi,\pi$) reference state, we obtain the acoustic and optical spin wave modes (as also mentioned in Ref.\cite{stratos2}) given by $\Omega_{ac}(k)=4J_1S\sqrt{(1-\lambda(1-\Gamma_k))^2-\gamma_k^2}$ and $\Omega_{op}(k)=4J_1S\sqrt{(1-\lambda(1+\Gamma_k))^2-\gamma_k^2+\Gamma_k}$.

For $\lambda>0.5$, similar calculation with $(\pi,0)$ order representing the zero-spin deviation state gives multiple acoustic modes with expressions $\Omega_1(k)=4J_1S\sqrt{(\lambda+\frac{1}{2}cos k_y)^2-(\lambda\Gamma_k+\frac{1}{2}cos k_x)^2}$ and $\Omega_2(k)=4J_1S\sqrt{(\lambda-\frac{1}{2}cos k_y)^2-(\lambda\Gamma_k-\frac{1}{2}cos k_x)^2}$.
\begin{figure}[htp]
\begin{center}
\epsfig{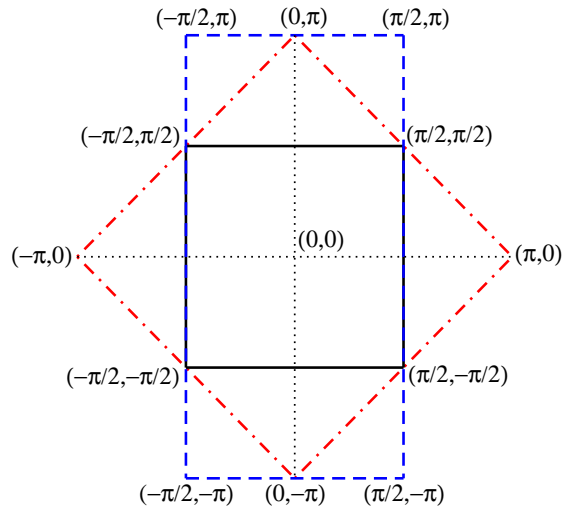}
\end{center}
\caption{(Color online) 1st Brillouin zone for $J_1-J_2$ model (black). Also shown are unit cell corresponding to $(\pi,\pi)$ (red, dash-dotted) and $(\pi,0)$ (blue, dashed) reference states.}
\label{figbz}
\end{figure}

Let's take a look at the Brillouin zones shown in Fig.\ref{figbz}. Here, the BZ for our 4-sublattice magnetic picture (let's call it BZ0), a square with corners at $(\pm\pi/2,\pm\pi/2)$, is shown (black lines) along with the unit cells BZ1 and BZ2, corresponding to $(\pi,\pi)$ (red, dash-dotted) and $(\pi,0)$ (blue, dashed) reference states respectively. For $J_2=0$, we have the BZ for the Neel AF, as shown by the square with dash-dotted lines in Fig.\ref{figbz}. The acoustic mode at $\Gamma$ point $k=(0,0)$ has zero dispersion energy as described by the Goldstone theorem. For $\lambda$ small, we have a stable $(\pi,\pi)$ reference state though the presence of $J_2$ term brings in 4 sublattices to the system and thus a further reduced BZ0. Due to band-folding, the $(\pi,0)$ point of BZ1 falls on the $\Gamma$ point of BZ0. And the dispersion energy for the acoustic mode at $k=(\pi,0)$ being non-zero, we get an band-folded optical mode with a non-zero spin gap at the zone center of the BZ. 

In the other limit of $J_1=0$, we have a classical ground state degeneracy between $(\pi,0)$ and $(0,\pi)$ states. For the $(\pi,0)$ spin reference state that we consider here, the $(0,\pi)$ vector of BZ2 folds back to the $\Gamma$ point due to band folding. But as the dispersion energy for the acoustic mode at $k=(0,\pi)$ is zero, we no more get an optical mode. Rather two distinct acoustic modes are obtained in the reduced picture.
These modes become degenerate in the limit $\lambda\rightarrow\infty$ where its expression becomes same as that of a $J_2$ antiferromagnet. It should be pointed out here that though the four sublattice picture gives the degeneracy between $(0,0)$ and $(0,\pi)$ states, they correspond to different points in the Brillouin zone and thus represent different spin wave modes. Our magnon dispersion expressions, when unfolded to the two-sublattice picture, matches with the dispersion relations reported in various literature\cite{thalmeier,erica,hamad,avinash1}. 

In this context, we should mention that for large frustration (as compared to small $\lambda$), the $Z_4$ lattice symmetry is broken\cite{chandra} and there is a degeneracy between $Q_1=(\pi,0)$ and $Q_2=(0,\pi)$ reference states. Within a two-sublattice picture with, say, $Q_1$ denoting the reference state, zero dispersions are found not only at the $\Gamma$ point but also for the $Q_2$ vector. A 4-sublattice description becomes more appropriate in this limit where zero magnon energies appear only at the $\Gamma$ point. Also the reduced BZ gives multiple magnon modes which in turn helps us understanding the multiple spin gaps observed experimentally in highly frustrated magnetic systems\cite{ewings,conceicao,zhao}. These systems, for example the iron pnictide compound $SrFe_2As_2$ (with typically $J_2\sim2J_1$), where uniaxial anisotropy is also present, witness multiple spin gaps in the neutron diffraction experiment (see Fig.3a and Fig.2a in Ref.\cite{zhao} and \cite{conceicao} respectively). A four-sublattice description of the $J_1-J_2$ AF recognizes the unit-cell doubling corresponding to the Hamiltonian and gives the multiple magnon branches to the system. These modes, in turn, get shifted differently in the presence of an anisotropy term leading to multiple spin gaps, as seen experimentally.  For example, our calculations can give rise to distinct spin gaps of $4S\sqrt{D^2+2D(J_2\pm J_1/2)}$ (for large $\lambda$) if an easy axis anisotropy term $-D\sum S_z^2$ is added to the Hamiltonian. Instead, a ferromagnetic term can also be added to break this $(0,0)-(0,\pi)$ mode degeneracy. Reference to the inelastic neutron scattering results from the iron pnictides can be made in this connection. They show maximum magnon energy at the ferromagnetic zone boundary, unlike a zero energy mode at the $(0,\pi)$ vector (as seen in this paper) of a $J_1-J_2$ antiferromagnet. A net ferro spin coupling in the ferro direction ($\hat y$ in this case) is thus required, in this case, to obtain the magnon dispersion in agreement with the experiments\cite{avinash3}.

\section{Hole dynamics} 
In order to incorporate hole motion in such a system we include NN hole-hopping. The resulting $t-J_1-J_2$ Hamiltonian is thus given by
\begin{eqnarray}
H_{tJ1J2}=H_{J1J2}-t\sum_{<i,j>,\sigma}(C_{i,\sigma}^\dagger C_{j,\sigma}+h.c.).
\label{eq5}
\end{eqnarray}
$C$ denoting the fermionic annihilation operators, maintaining the constraint of no double occupancy. The $t$ term can be linearized after rewriting it in terms of spin wave operators and the slave fermions as worked out in Ref.\cite{liu} for a 2D $t-J$ model. The only difference in our case will be due to having 4 sublattices and a Green's function matrix that has non-zero off-diagonal entries.

Let us consider $(\pi,\pi)$ reference state for small $\lambda$. In this case we have coupled Dyson's equations for a $4\times4$ Green's function matrix:
\begin{eqnarray}
G^{(n)}_{ij}=G^{(0)}_{ij}+\sum_kG^{(0)}_{ii}\Sigma^{(n)}_{ik}G^{(n)}_{kj}\nonumber\\
i.e.,~~~ \sum_{k=1}^4(\delta_{ik}-G^{(0)}_{ii}\Sigma^{(n)}_{ik})G^{(n)}_{kj}=G^{(0)}_{ij}
\label{eq6}
\end{eqnarray}
where the superscripts indicate the iteration numbers.
By symmetry, all $G_{ii}$'s are same and off-diagonal $G_{ij}$'s are non-zero only if $(i,j)$ pairs are (1,3) or (2,4). Solving matrix equation gives
\begin{eqnarray}
G_{11}&=&\frac{G^{(0)}_{11}(1-G^{(0)}_{11}\Sigma_{11})}{(1-G^{(0)}_{11}\Sigma_{11})^2-(G^{(0)}_{11}\Sigma_{13})^2}\nonumber\\
G_{13}&=&\frac{(G^{(0)}_{11})^2\Sigma_{13}}{(1-G^{(0)}_{11}\Sigma_{11})^2-(G^{(0)}_{11}\Sigma_{13})^2}
\label{eq8}
\end{eqnarray}
Here argument of $(k,\omega)$ are implied for all the variables.

 Within the non-crossing approximation (NCA), hole self energy expressions are obtained as
\begin{align}
\Sigma_{11}(k,\omega)=&c_0[(g_{12j}^2(k,q)+f_{14j}^2(k,q))G_{11}(k-q,\omega-\Omega_{j,q})\nonumber\\&+2g_{12j}(k,q)f_{14j}(k,q)G_{13}(k-q,\omega-\Omega_{j,q})],\nonumber\\
\Sigma_{13}(k,\omega)=&c_0[(g_{12j}^2(k,q)+f_{14j}^2(k,q))G_{13}(k-q,\omega-\Omega_{j,q})\nonumber\\&+2g_{12j}(k,q)f_{14j}(k,q)G_{11}(k-q,\omega-\Omega_{j,q})].
\label{eq9}
\end{align}
where sum over ${\bf q}$, $j$ (1 and 3) and $j^\prime$ (2 and 4) are implied. $c_0=\frac{16}{N}t^2$, $g_{ijl}({\bf k,q})=(U_{jl}({\bf q})cos(k_x)+U_{il}({\bf q})cos(k_x-q_x))$ and $f_{ijl}({\bf k,q})=(U_{jl}({\bf q})cos(k_y)+U_{il}({\bf q})cos(k_y-q_y))$. $\Omega_{j,q}$ denotes the $j$-th magnon mode energy at wave vector $q$.

\begin{figure}[htp]
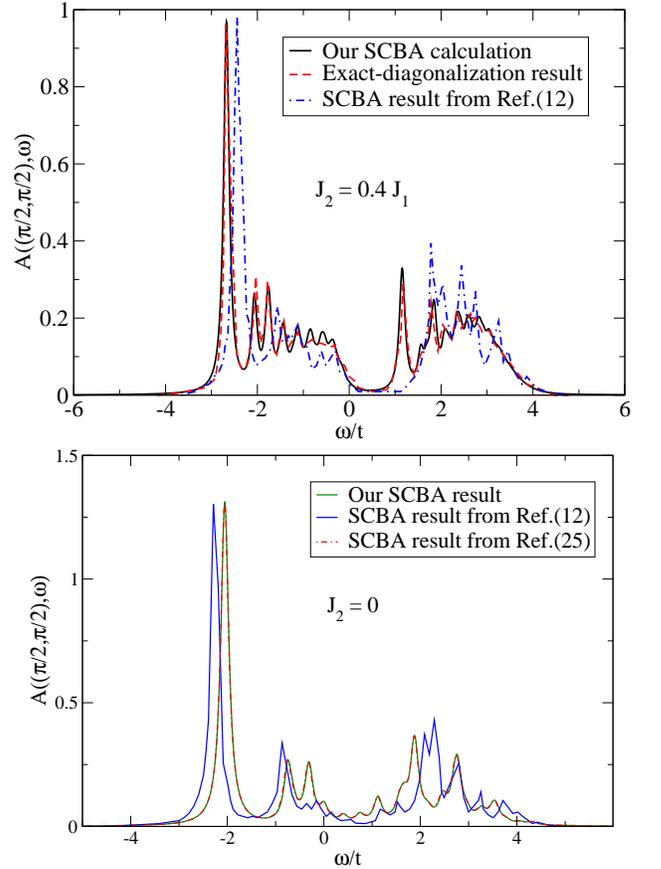

\begin{center}
\begin{tabular}{c}
\epsfig{file=spectra-compare.eps,width=.95\linewidth,clip=}
\\
\epsfig{file=spectra-compare-b.eps,width=.9\linewidth,clip=}
\end{tabular}
\end{center}
\caption{(Color online) ($\pi/2,\pi/2$) spectra of a $4\times4$ lattice for $J_1=0.4t$.}
\label{fig3}
\end{figure}

With this self-energy expression we solve the Dyson's equation (Eq.\ref{eq6}) iteratively following SCBA to obtain the single hole spectra which is given as $A(k,\omega)=-Img[G(k,\omega)]/\pi$. Here $G(k,\omega)=G_{11}(k,\omega)\pm G_{13}(k,\omega)$ denotes the eigenvalues of the Green's function matrix of which $G_{11}(k,\omega)+G_{13}(k,\omega)$ corresponds to the $k$ value in the 1st BZ whereas $G_{11}(k,\omega)-G_{13}(k,\omega)$ corresponds to the other mode appearing due to folding of the bands in the reduced BZ of the four-sublattice structure.

Our findings show that the general feature of the hole spectra is to have a low energy quasi-particle like excitation followed by a series of higher energy excitations called string states\cite{liu}. The lowest quasi-particle (QP) peak is found at $(\frac{\pi}{2},\frac{\pi}{2})$ and as we increase $\lambda$ from 0 to 0.4, the $(\frac{\pi}{2},\frac{\pi}{2})$ peak gradually shifts to lower energy and reduces its strength.

\begin{figure}[htp]
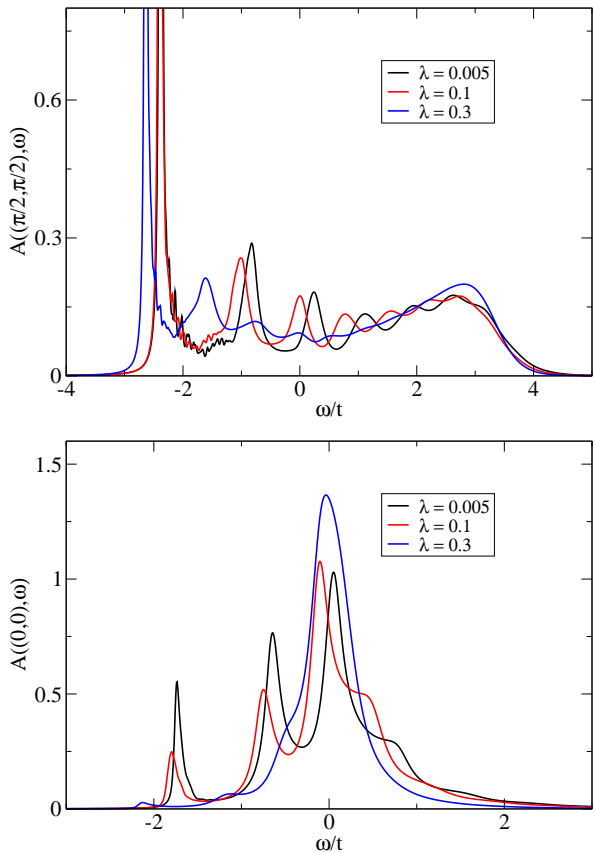

\begin{center}
\epsfig{file=figure4a.eps,width=.9\linewidth,clip=}\vskip .1 in
\epsfig{file=figure4b.eps,width=.9\linewidth,clip=}
\end{center}
\caption{(Color online) Hole spectral function $A(k,\omega)$ for $k=(\pi/2,\pi/2)$ (top) and $k=(0,0)$ (bottom) for different small values of $\lambda$ at $J_1=0.3t$.}
\label{fig4}
\end{figure}

\begin{figure}[htp]
\begin{center}
\epsfig{file=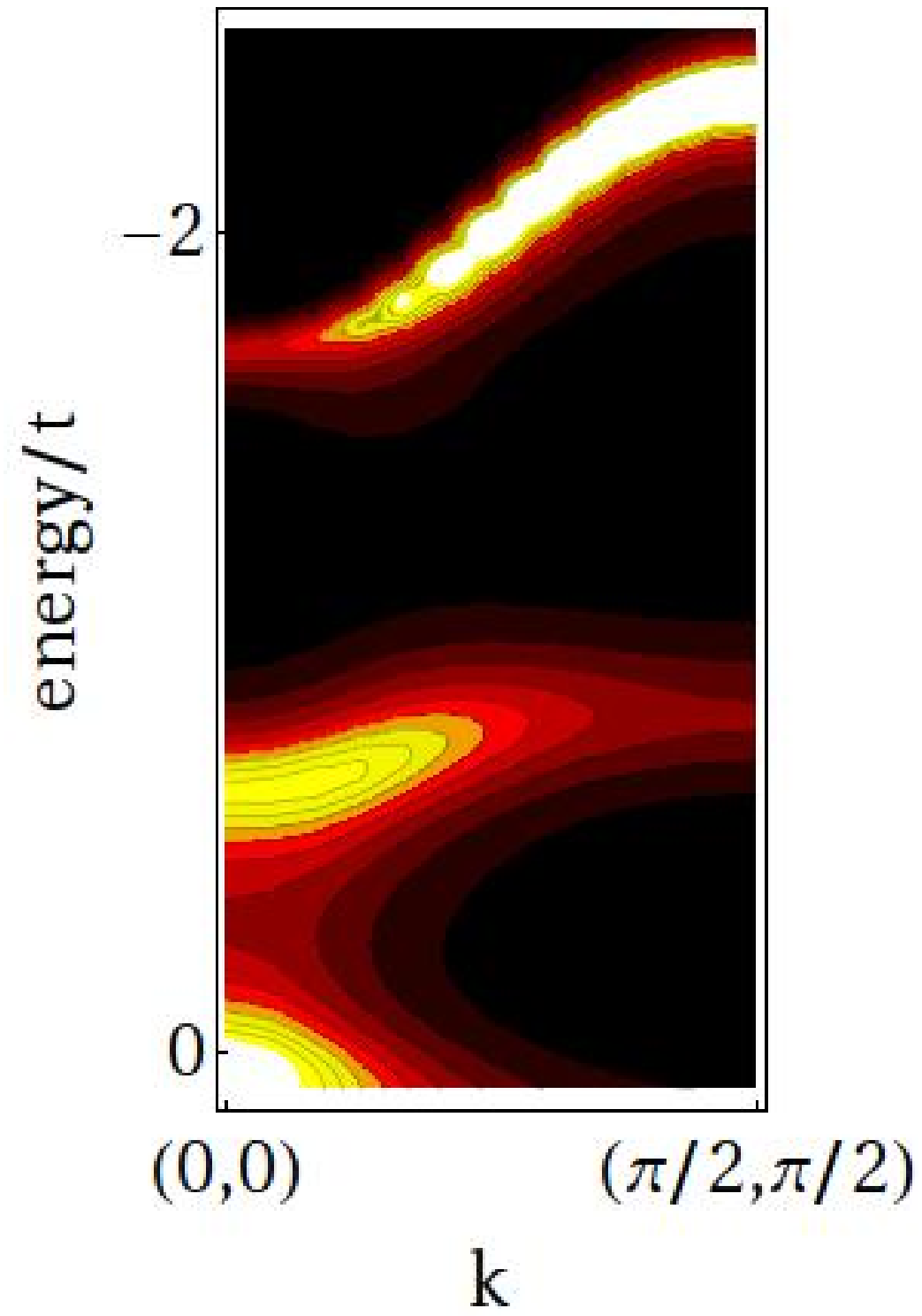,width=.48\linewidth,clip=}\hskip .1 in
\epsfig{file=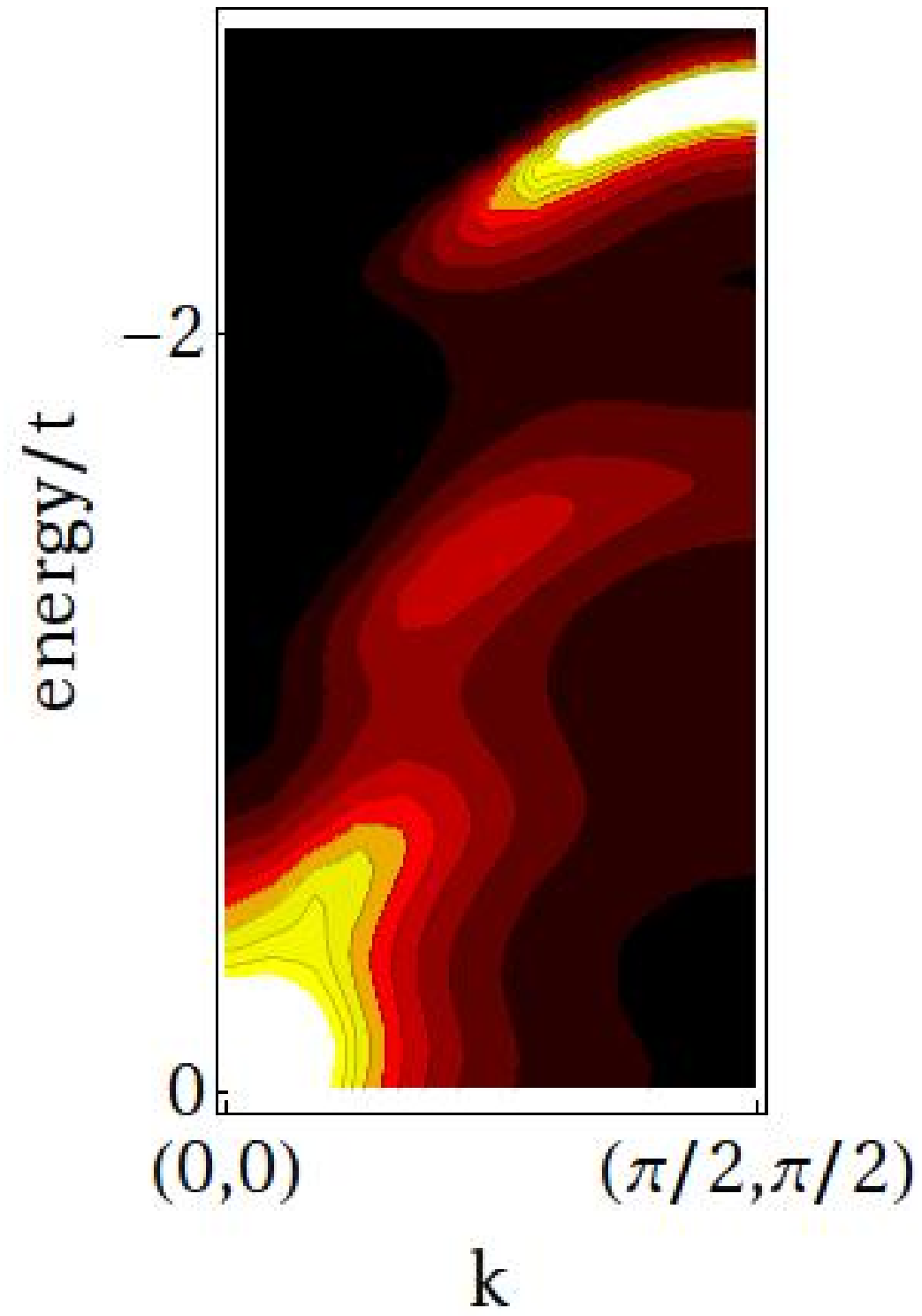,width=.48\linewidth,clip=} \vskip .05 in
\epsfig{file=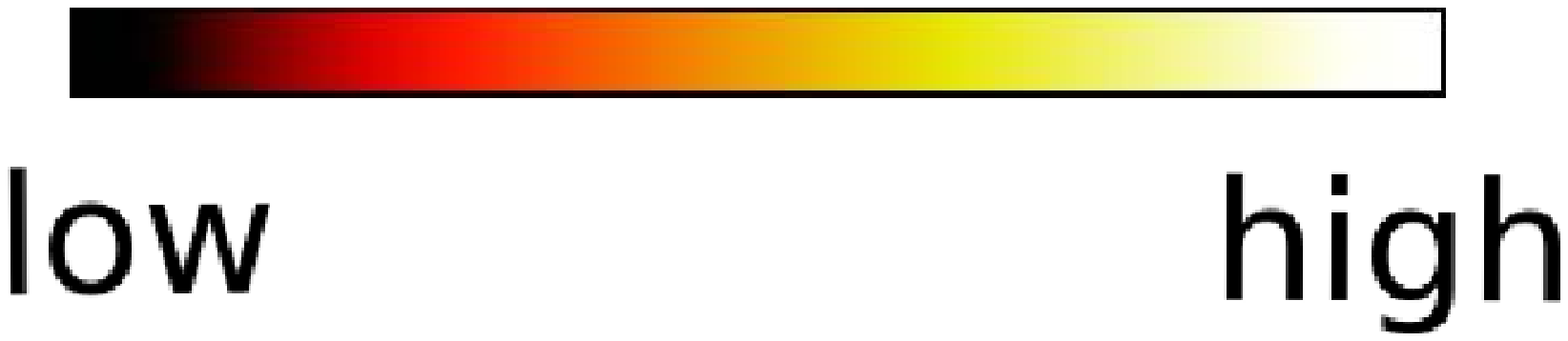,angle=-90,width=.3\linewidth,clip=}
\end{center}
\caption{(Color online) Spectral intensity along the nodal direction from $(0,0)$ to $(\pi/2,\pi/2)$ points in $k$ space in a $48\times48$ lattice for $\lambda=0.005$ (left) and $\lambda=0.3$ (right) respectively.}
\label{fign}
\end{figure}

\begin{figure}[htp]
\begin{center}
\epsfig{file=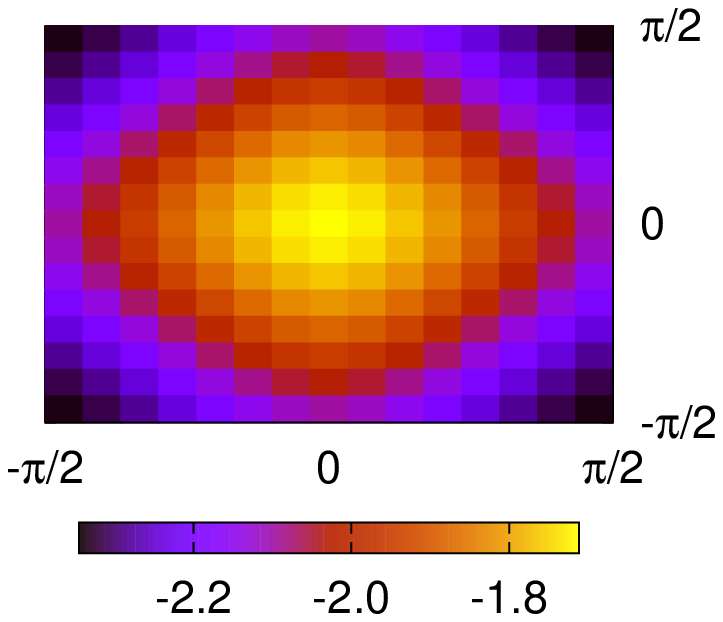,width=.48\linewidth,clip=}
\epsfig{file=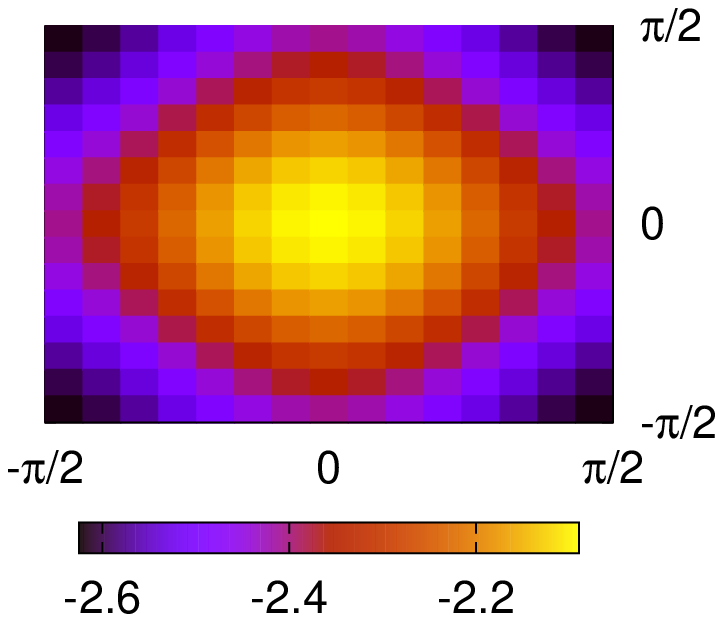,width=.48\linewidth,clip=} 
\end{center}
\caption{(Color online) Surface plots for QP Hole energies in the 1st BZ in a $32\times32$ lattice for $\lambda=0.005$ (left) and $\lambda=0.3$ (right) respectively.}
\label{splot}
\end{figure}

Our spectral results in the $J_2\rightarrow 0$ limit are essentially the same as obtained from SCBA calculations done on a $t-J$ model\cite{liu}. The comparison at non-zero values of $J_2$ has also been made. In Fig.\ref{fig3} top-panel, we show the $(\pi/2,\pi,2)$ spectra of a $4\times4$ lattice at $\lambda=0.4$ (with $J_1=0.4t$) obtained by our SCBA calculation with exact diagonalization (ED) plot taken from Ref.\cite{hamad} demonstrating good match between the two. However we find that our results are not identical to the SCBA results of Ref.\cite{hamad} also shown in Fig.\ref{fig3}, though our results match much better with the exact result both in position and relative-strengths of the spectral peaks - corresponding to both the lowest energy QP and the high energy string excitations. As a further check, we also plot our results as well as SCBA results shown in Ref.\cite{hamad,liu} together in Fig.\ref{fig3} bottom-panel. The comparison clearly shows that our plot falls essentially on top of that from Ref.\cite{liu}, but shows discernible difference with the plot from Ref.\cite{hamad} (see note \cite{note} in the references).

In Fig.\ref{fig4}, we have shown the hole spectra for $\lambda$= 0.005, 0.1 and 0.3 respectively ($J_1=0.3t$) obtained from a $32\times32$ lattice for $k=(\pi/2,\pi/2)$ and $k=(0,0)$. The energy resolution of our numerical calculation is $\Delta\omega=0.005t$ and the small parameter $\eta=0.02t$. With these specifications, $32\times32$ lattice represent a large enough system to avoid the finite size effects in the spectral plots\cite{liu}. A hole in a 2D $t-J$ model prefers to move with wave vector $(\pi/2,\pi/2)$. We observe that with NNN AF interaction present, this remains the case for $\lambda\leq0.4$. In the $t-J_1-J_2$ model, with $J_2$ switched on starting from zero, low energy spectral peaks move to lower energies and this shift becomes more for higher values of $J_2$. Peaks along $(\pi/2,\pi/2)$ to $(\pi/2,0)$ directions which were already close to the lowest peak at $(\pi/2,\pi/2)$ for $\lambda=0$, becomes even closer as $\lambda$ is increased. Nevertheless the $(\pi/2,\pi/2)$ peak stays the lowest energy peak as long as $\lambda\leq0.4$. We also notice loss of intensity in the QP as well as the string states as $J_2$ is increased gradually. This enables a smooth transfer of spectral intensity from lowest peak at $(\pi/2,\pi,2)$ to the highest peak at $(0,0)$ resembling a water-fall as shown in Fig.\ref{fign}. This effect of frustration (as long as the $(\pi,\pi)$ reference state remains stable) is, thus, the same as that of electron-phonon coupling\cite{kar1} or doping\cite{kar2} in smearing out the long lived well-defined QP excitations and thereby producing smooth transitions in spectral intensity from the low energy peak to the high energy peak. This so called 'waterfall'-feature in intensity is observed along the nodal direction in the ARPES results of the superconducting cuprates\cite{damascelli}. The comparison can even be improved if the finiteness of the onsite coulomb repulsion between hole/electrons is considered as well ($e.g.$, see Ref.\cite{avinash2} for calculations using a Hubbard model with extended hoppings).
\begin{figure}[htp]
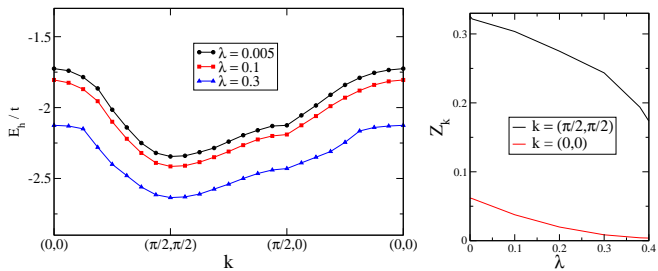

\begin{center}
\epsfig{file=figure5.eps,width=.63\linewidth,clip=}\hskip .06 in
\epsfig{file=residue2.eps,width=.35\linewidth,clip=}
\end{center}
\caption{(Color online) Left: Hole QP energy band along the boundary of irreducible BZ. Right: Decrease of QP residue with $\lambda$.}
\label{fig5}
\end{figure}

Fig.\ref{splot} shows the QP hole energies ($E_h$) in the 1st Brillouin zone for $\lambda=0.005$ and $0.3$. The lowest energy excitations are created around $(\pm\pi/2,\pm\pi/2$) points. These are the most likely hole excitations to form during photo-emission from the undoped AF samples producing hole-pockets around $(\pm\pi/2,\pm\pi/2$) points. On the other hand, the largest $E_h$ values are obtained at and around the $\Gamma$ point. Though the features for hole bands do not show any visible change between $\lambda=0.005$ and $0.3$, the energy shifts are clearly seen between the two frustration levels.
Also in Fig.\ref{fig5} left panel, we have constructed the hole energy band along the boundary of the irreducible BZ, namely $(0,0)\rightarrow(\pi/2,\pi/2)\rightarrow(\pi/2,0)\rightarrow(0,0)$ for similar values of $\lambda$. It clearly shows how the QP peak positions lower in energy with an increase in $J_2$. At the same time the hole bandwidth $W$ also decreases as $W\sim2(J_1-J_2)$. 

The construction of this hole-bands involves defining the lowest energy excitations for different $k$-values to be the QP excitations of the holes. But calculation of the residues ($Z_k$) of those lowest energy peaks shows gradual decrease\cite{hamad} in value (see Fig.\ref{fig5} right panel) as $\lambda$ is increased from zero. We see that, for $\lambda\sim0.4$, the residue for the $(0,0)$ (or $(\pi,\pi)$) vector almost vanishes. So we find that the QP description of the hole excitations, for such $k$ vector, does not remain valid even for such moderate value of frustration.

For large frustration, with a $(\pi,0)$ reference state, neighboring spins along $y$-directions prefer to align ferromagnetically while that along $x$ prefers an AF order. 
The hopping term in the Hamiltonian gives the on-site hole energies to be $\epsilon^{(0)}(k)=2t$cos($k_y$) whereas a typical off-diagonal term there is of the form $h_{i,k}^\dagger h_{j,k-q}\alpha_{m,q}$, $h_{i,k}$ being the hole annihilation operator of $i$-th type at wave-vector $k$. The vector $k=(0,\pi)$ falls within the 1st Brillouin zone of the two-sublatice picture (the dashed-line rectangle in Fig.\ref{figbz}) and it has zero magnon energies ($\Omega_{m,(0,\pi)}=0$ for all $m$). Thus the degenerate magnon modes $k=(0,0)$ and $(0,\pi)$ contribute different hole energies due to the NN hole-hoppings. 
Also, $q=(0,\pi)$ in the term $h_{i,k}^\dagger h_{j,k-q}\alpha_{m,q}$ implies that a hole, by consecutive annihilation-creation, can change its wave-vector by $q$ (having non-zero magnitude), yet not changing its energy at all. 
In the four-sublattice picture, $(0,\pi)$ vector is no more within the 1st Brillouin zone. 
But here the degeneracy among the magnon modes increases manifold. These cases are dealt with proper linear combination of eigen functions so that the coefficients of the Bogoliubov transformation are obtained maintaining the bosonic commutation relations. As far as the hole spectral functions are concerned, the holes in a $(\pi,0)$ (or a $(0,\pi)$) background, with a NN hopping term, always prefers to move along the ferromagnetic alignment of spins (the $y$-direction for a $(\pi,0)$ reference state) in the lattice, as moving in the orthogonal direction involves leaving behind strings of anti-aligned spins\cite{liu} in the expense of high energy cost. That is why, string excitations are unlikely to form among the low energy holes. The hole spectral calculations for large frustration, owing to these special features, need thorough analysis and we would like to discuss this matter in detail later in a future communication.

\section{Summary and Discussion}
In this paper, we have studied the spin dynamics of a 2D $J_1-J_2$ model and found the acoustic and optical magnon modes about the state $k=(\pi,\pi)$. Contrarily, two acoustic and no optical modes are obtained when large values of $\lambda$ are considered and $k=(\pi,0)$ is used as the reference state.

Hole spectra are obtained from a 2D $t-J_1-J_2$ model for small $\lambda$. We observe the QP and string-like excitations in the spectral behavior. The energy shift and bandwidth reduction for the hole QP is also witnessed as $\lambda$ is increased from zero.

Our calculations, in general, can represent a nice exercise to show how the magnon modes can be studied for a superlattice structure starting from suitable reference states and how the hole spectra can be obtained for the same from the eigenvalues of the Green's function matrix, obtained by solving the Dyson's equation self-consistently. 

As we said earlier, we plan to present this hole spectral analysis in the largely frustrating limit rigorously later in a future communication - which can also include special situations with the reference state being a superposition of the degenerate magnetic orders\cite{caretta,hamad2}. Our four-sublattice calculations can also stand as an important building block for performing the more involved hole spectral calculations in the multi-orbital $t-J_1-J_2$ models\cite{stratos2,yu} in the limit of large frustration (typically $\lambda\sim2$), so that comparison with the ARPES results obtained from FeAs superconduting materials\cite{xia,nakayama} becomes possible.

\section*{Acknowledgement}
The author benefited from discussions with Dr. Erica Carlson on magnon dispersions in superlattices. Acknowledgement also goes to Saha Institute of Nuclear Physics, India for the funding support.


\begin{thebibliography}{99}
\bibitem{note} We should mention here that our SCBA calculations using only 2 sublattices matches with our results with 4 sublattices presented in this paper. And they match with exact results as demonstrated in Fig.\ref{fig3}, much better than that obtained by SCBA calculations of Ref.\cite{hamad}. We would like to add at this point that the expression for hole-magnon coupling coefficient used in Ref.\cite{hamad}, in the $J_2=0$ limit, would be different from that used by several groups as can be found in Ref.\cite{liu,kane,marsiglio} and the follow-up cited works. Also the self energy expression of page 3 in Ref.\cite{hamad} is not correct (with $\omega_{k-q}$ in place of $\omega_{q}$).
\end{thebibliography}
\end{document}